\documentclass[submission, Phys]{SciPost}
\usepackage{xcolor}
\definecolor{scipostblue}{RGB}{0,51,102}
\hypersetup{
    colorlinks,
    linkcolor={red!50!black},
    citecolor={blue!50!black},
    urlcolor={blue!80!black}
}

\usepackage[normalem]{ulem}
\usepackage[bitstream-charter]{mathdesign}
\usepackage{subcaption}
\usepackage{booktabs}

\DeclareSymbolFont{usualmathcal}{OMS}{cmsy}{m}{n}
\DeclareSymbolFontAlphabet{\mathcal}{usualmathcal}

\fancypagestyle{SPstyle}{
\fancyhf{}
\lhead{\colorbox{scipostblue}{\bf \color{white} ~SciPost Physics Community Reports }}
\rhead{{\bf \color{scipostdeepblue} ~Submission }}

\fancyfoot[C]{\textbf{\thepage}}
}

\usepackage{array}
\newcommand{\df}{{\rm d}}

\newcommand{\muf}{\mu_F}
\newcommand{\mur}{\mu_R}

\newcommand{\nn}{\nonumber\\}

\newcommand{\fig}[1]{Fig.\ \ref{#1}}
\newcommand{\tab}[1]{Table\ \ref{#1}}
\newcommand{\sect}[1]{Section\ \ref{#1}}

\begin{document}

\pagestyle{SPstyle}


{\flushright{P3H-25-079, TTK-25-24, IPPP/25/69}}
\begin{center}{\Large \textbf{\color{scipostdeepblue}{
Threshold improved $Z H$ production at the LHC\\
}}}\end{center}
\begin{center}
\textbf{
\begin{center}
	Goutam Das\textsuperscript{1$\star$},
	Chinmoy Dey\textsuperscript{2,3$\dagger$},
	M. C. Kumar\textsuperscript{2$\ddagger$}, and
	Kajal Samanta\textsuperscript{4$\circ$}
\end{center}
}
\end{center}

\begin{center}
\begin{center}
	{\bf 1} Institut f{\"u}r Theoretische Teilchenphysik und Kosmologie, RWTH Aachen University, \\D-52056 Aachen, Germany
	\\
	{\bf 2} Department of Physics, Indian Institute of Technology Guwahati,\\ Guwahati-781039, Assam, India
	\\
	{\bf 3} Theoretical Physics Division, Physical Research Laboratory,\\ Navrangpura, Ahmedabad 380009, India
	\\
	{\bf 4} Institute for Particle Physics Phenomenology, Durham University,\\ Durham DH1 3LE, United Kingdom
\end{center}

$\star$~\href{mailto:goutam@physik.rwth-aachen.de}{\small goutam@physik.rwth-aachen.de}\,,
$\dagger$~\href{mailto:d.chinmoy@iitg.ac.in}{\small d.chinmoy@iitg.ac.in}\,,
$\ddagger$~\href{mailto:mckumar@iitg.ac.in}{\small mckumar@iitg.ac.in}\,,
$\circ$~\href{mailto:kajal.samanta@durham.ac.uk}{\small kajal.samanta@durham.ac.uk}\,
\end{center}

\section*{\color{scipostdeepblue}{Abstract}}
\textbf{\boldmath{%
We present precise theoretical results for the $ZH$ production cross section and
invariant mass distribution at the Large Hadron Collider (LHC) taking into account the effects of soft gluons. We improve both quark-initiated and gluon-initiated subprocesses through threshold resummation within the QCD framework and present combined results relevant for $13.6$ TeV LHC.  
}
}
\vspace{\baselineskip}

\noindent\textcolor{white!90!black}{%
\fbox{\parbox{0.975\linewidth}{%
\textcolor{white!40!black}{\begin{tabular}{lr}%
  \begin{minipage}{0.6\textwidth}%
    {\small Copyright attribution to authors. \newline
    This work is a submission to SciPost Phys. Comm. Rep. \newline
    License information to appear upon publication. \newline
    Publication information to appear upon publication.}
  \end{minipage} & \begin{minipage}{0.4\textwidth}
    {\small Received Date \newline Accepted Date \newline Published Date}%
  \end{minipage}
\end{tabular}}
}}
}

\nolinenumbers

\vspace{10pt}
\noindent\rule{\textwidth}{1pt}
\tableofcontents
\noindent\rule{\textwidth}{1pt}
\vspace{10pt}


\section{Introduction}
\label{sec:intro}
The associated production of the Higgs boson with a $Z$ boson at the Large Hadron Collider (LHC) plays a crucial role in probing the Higgs couplings to the electroweak gauge bosons. 
Both the ATLAS and CMS collaborations are actively pursuing precision measurements of this process \cite{ATLAS:2020fcp, CMS:2024tdk, CMS:2024fkb,  ATLAS:2024yzu}. 
To fully exploit the experimental precision, equally precise theoretical predictions are essential. The dominant production mode for $Z H$ in LHC is through the Drell-Yan (DY) type quark-antiquark annihilation $(q\bar{q} \rightarrow Z^* \rightarrow ZH)$ for which the higher order Quantum Chromodynamics (QCD) correction has been known to next-to-next-to-next-to-leading order (N3LO) \cite{Baglio:2022wzu} accuracy. In addition to the DY-type contribution, other subprocesses also contribute to $Z H$ production, such as bottom quark annihilation \cite{Ahmed:2019udm} and top quark loop–induced contribution \cite{Brein:2011vx}. However, these contributions are subleading, typically contributing at the sub-percent level relative to the NLO QCD DY-type prediction. Electroweak (EW) corrections to the DY-type channel have also been computed at NLO \cite{Ciccolini:2003jy}, yielding a negative correction of approximately $5\%$ relative to the NLO QCD result. Starting from N2LO, the $Z H$ production receives contributions from the gluon fusion channel. Although this subprocess is suppressed by two powers of the strong coupling constant ($\alpha_S$) relative to the DY-type channel, the suppression is largely offset by the substantial gluon luminosity at the LHC, leading to a significant overall contribution.

Given its growing importance, the gluon fusion subprocess ($ggZH$) has been extensively studied in the literature at both leading order (LO) \cite{Dicus:1988yh, Kniehl:1990iva, Kniehl:1990zu, Kniehl:2011aa} and next-to-leading order (NLO) \cite{Altenkamp:2012sx, Hasselhuhn:2016rqt, Davies:2020drs, Alasfar:2021ppe, Bellafronte:2022jmo, Degrassi:2022mro, Wang:2021rxu, Chen:2020gae, Chen:2022rua}. At NLO, the total cross section of this channel roughly doubles compared to the prediction of LO, while the theoretical uncertainty due to renormalization and factorization scale variations is reduced to about $15\%$. The fixed order results thus suffer from the large threshold logarithms arising from soft gluons emission. 
Indeed, the threshold soft-virtual (SV)  logarithms at NLO can contribute to $90-99\%$ of the complete NLO results in the range $Q=350-2000$ GeV \footnote{Similar observation has also been made for the DY\cite{Ahmed:2014cla} and Higgs \cite{Catani:2001ic} processes.}.
By resumming these SV logarithms to all orders, one obtains predictions that are stable and well-behaved across the relevant kinematic regions.
%
The formalism for threshold resummation is well established in the literature \cite{Sterman:1986aj, Catani:1989ne, Catani:1990rp, Kidonakis:1997gm, Moch:2005ba, Laenen:2005uz, Kidonakis:2005kz, Ravindran:2005vv, Ravindran:2006cg, Idilbi:2006dg, Becher:2006mr, Ahmed:2020nci}, 
and has been successfully applied to improve theoretical predictions for both inclusive cross-sections and differential observables such as invariant mass distributions.

In this report, we present an improved theoretical description of the $Z H$ process at LHC by incorporating threshold resummation effects at the SV level for gluon fusion  as well as for the DY-type channels to $Z H$ production. Our analysis \cite{Das:2025wbj} covers both the total cross-section and the invariant mass distribution of the $Z H$ pair. We employ the Born-improved gluon fusion framework, which has proven effective in the case of inclusive Higgs production, and we expect it to yield similarly reliable results for the $Z H$ channel. The article is organized as follows: In \sect{sec:theory}, we introduce the key theoretical formulas. In \sect{sec:results}, we provide a phenomenological study for the gluon fusion subprocess, combining it with DY-type contributions to present complete results for $pp$ collisions with ${\cal O}(\alpha_s^3)$ accuracy.  Finally, we conclude in \sect{sec:conclusion}.

\section{$Z H$ production at the LHC}
\label{sec:theory}
The hadronic cross-section for $Z H$ production at LHC can be written as,

\begin{align}\label{eq:had-xsect}
Q^2 \frac{\df \sigma}{\df Q^2}  
=  
\sum_{a,b}\int_0^1 \df x_1\int_0^1 \df x_2 \,\,f_{a}(x_1,\mu_F^2)\,
f_{b}(x_2,\mu_F^2) 
\int_0^1 \df z~ \delta \left(\tau-zx_1 x_2\right)
Q^2 \frac{\df \widehat\sigma_{ab}(z,\muf^2)}{\df Q^2} \, ,
\end{align}
where $f_{a,b}$ are the parton distribution functions (PDFs) for parton $a,b$ in the incoming protons
and $\widehat\sigma_{ab}$ is the  partonic coefficient function.
The hadronic and partonic threshold variables $\tau=Q^2/S$ 
and $ z= Q^2/\widehat{s}$ are defined in terms of respective center-of-mass energies $S$ and $\widehat{s}$. Here $Q$ is 
the invariant mass of the $Z H$ system and $\muf$ is the factorization scale.
The partonic coefficient function can be decomposed in soft-virtual ($\Delta_{ab}^{\rm SV}$) and regular ($\Delta_{ab}^{\rm REG}$) parts
at each order in $\alpha_S$.
The SV part contains all dominant singular contribution in the limit $z \to 1$ whereas the regular part contains sub-dominant contributions. 
The SV logarithms can be resummed in Mellin-$N$ space and corresponding resummed partonic coefficient takes the form,
\begin{align}\label{eq:resum-partonic}
	\frac{1}{\widehat{\sigma}^{(0)}_{ab}(Q^2)}Q^2\frac{\df \widehat{\sigma}_{N,ab}^{\rm N{\it n}LL}}{\df Q^2}
	= \int_0^1 \df z ~ z^{N-1}   \Delta^{\rm SV}_{ab}(z) 
	\equiv g_{0}(Q^2) \exp \left( G^{\rm SV}_N \right)\,.
\end{align}
The function $G^{\rm SV}_N$ contains the universal threshold exponent and determines the resummed accuracy through its expansion (see for example in \cite{Das:2019btv,AH:2020cok,Das:2022zie}). The constant $g_{0}(Q^2)$ contains the process-dependent information (see \cite{Das:2022zie,Das:2025wbj}). Using \textit{minimal prescription} \cite{Catani:1996yz} for Mellin inversion, one can finally find resummed results in the physical $z$-space which can be also matched to the corresponding fixed order results to incorporate missing regular corrections,
\begin{align}\label{eq:MATCHING}
	Q^2\frac{\df \sigma^{\rm N{\it n}LO+N{\it n}LL}_{ab}}{\df Q^2}
	=&
	Q^2 \frac{\df {\sigma}^{\rm N{\it n}LO}_{ab} }{\df Q^2}
	+
        \sum_{ab \in \{gg, q\bar{q}\} }
        \widehat{\sigma}^{(0)}_{ab}(Q^2)
	\int_{c-i\infty}^{c+i\infty}
	\frac{\df N}{2\pi i}
	\tau^{-N}
	f_{a,N}(\muf)
	f_{b,N}(\muf)
	\nn
	&\times 
	\Bigg( 
		Q^2\frac{\df \widehat{\sigma}_{N,ab}^{\rm N{\it n}LL}}{\df Q^2} 
		- 	
		Q^2\frac{\df \widehat{\sigma}_{N,ab}^{\rm N{\it n}LL}}{\df Q^2} \Bigg|_{\rm tr}
	\Bigg) \,.
\end{align}
Other prescriptions, for instance the Borel prescription \cite{Forte:2006mi, Abbate:2007qv, Bonvini:2010tp}, may lead to differences with respect to the minimal prescription that are confined to subleading terms.
\section{Results}
\label{sec:results}
All the ingredients necessary for performing soft-gluon resummation in both the gluon fusion and Drell–Yan (DY)–type subprocesses for $Z H$ production are available in \cite{Das:2022zie,Das:2025wbj}, and can be used to quantitatively assess their impact at $13.6$ TeV LHC. Our choice of parameters are given below:
\begin{align}\label{eq:parametrs}
	\sqrt{S} = 13.6 \text{ TeV}, \quad \text{PDF} = {\tt PDF4LHC21\_40} \text{\cite{PDF4LHCWorkingGroup:2022cjn}}, \quad \alpha_S (m_Z) = 0.1180\,, \nn
    \alpha \simeq 1/127.93, \quad m_Z = 91.1880 \text{ GeV}, \quad \Gamma_Z = 2.4955 \text{ GeV}\,, \nn
    m_W = 80.3692 \text{ GeV}, \quad m_t = 172.57 \text{ GeV}, \quad m_H = 125.2 \text{ GeV}\,.
\end{align}
The weak mixing angle is then determined by $\text{sin}^2\theta_\text{w} = (1 - m_W^2/m_Z^2)$ corresponding to the Fermi constant $G_F \simeq 1.2043993808\times 10^{-5} \text{ GeV}^{-2}$. 
To account for different types of uncertainties, we have used the following formulas;
\begin{subequations}
\begin{alignat}{3} \label{eq:uncertainty}
\delta(\text{PDF}) &=  \left(\sum_{i=1}^{40} (\sigma(i) - \sigma(0) )^2   \right)^{1/2} \,,
\\
    \delta(\alpha_s) 
   & =
    \left|\frac{3}{4}
    \frac{\sigma(\alpha^c_s(m_Z)=0.119) - \sigma(\alpha^c_s(m_Z)=0.117)}{\sigma(\alpha^c_s(m_Z)=0.118)} \right| ,
\\
\delta(\text{PDF} + \alpha_s) &= \sqrt{\delta( \alpha_s)^2 + \delta(\text{PDF})^2} \,.
\end{alignat}
\end{subequations}
For the scale uncertainty, we use seven-point scale variation where we vary both $\mu_R$ and $\mu_F$  by factor $2$ or $1/2$ around the central value with constraints $|\ln (\mu_R/\mu_F)| \leq \ln(2)$.

\begin{figure}[ht!]
\centerline{
\includegraphics[scale=0.35]{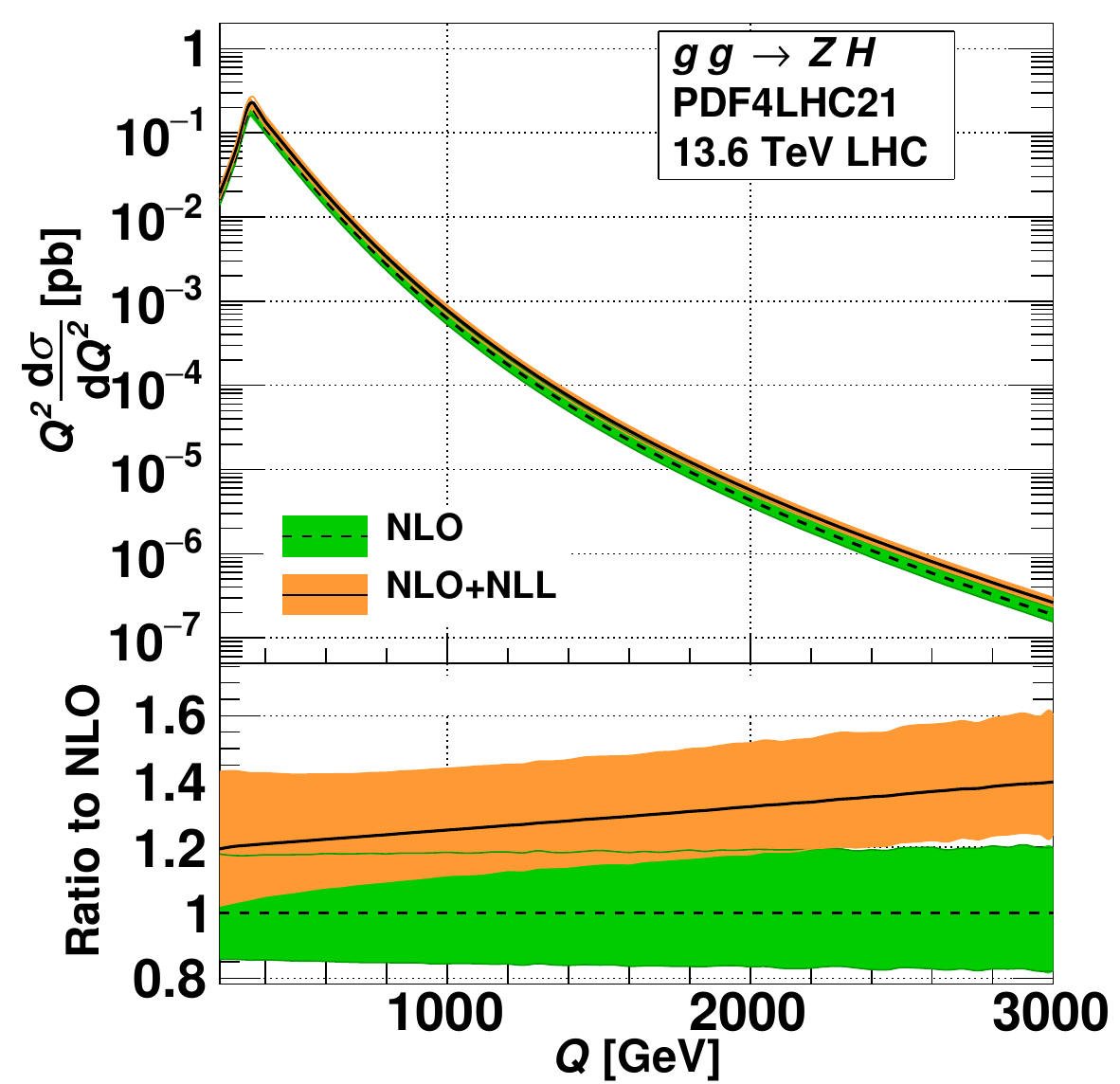}
\includegraphics[scale=0.36]{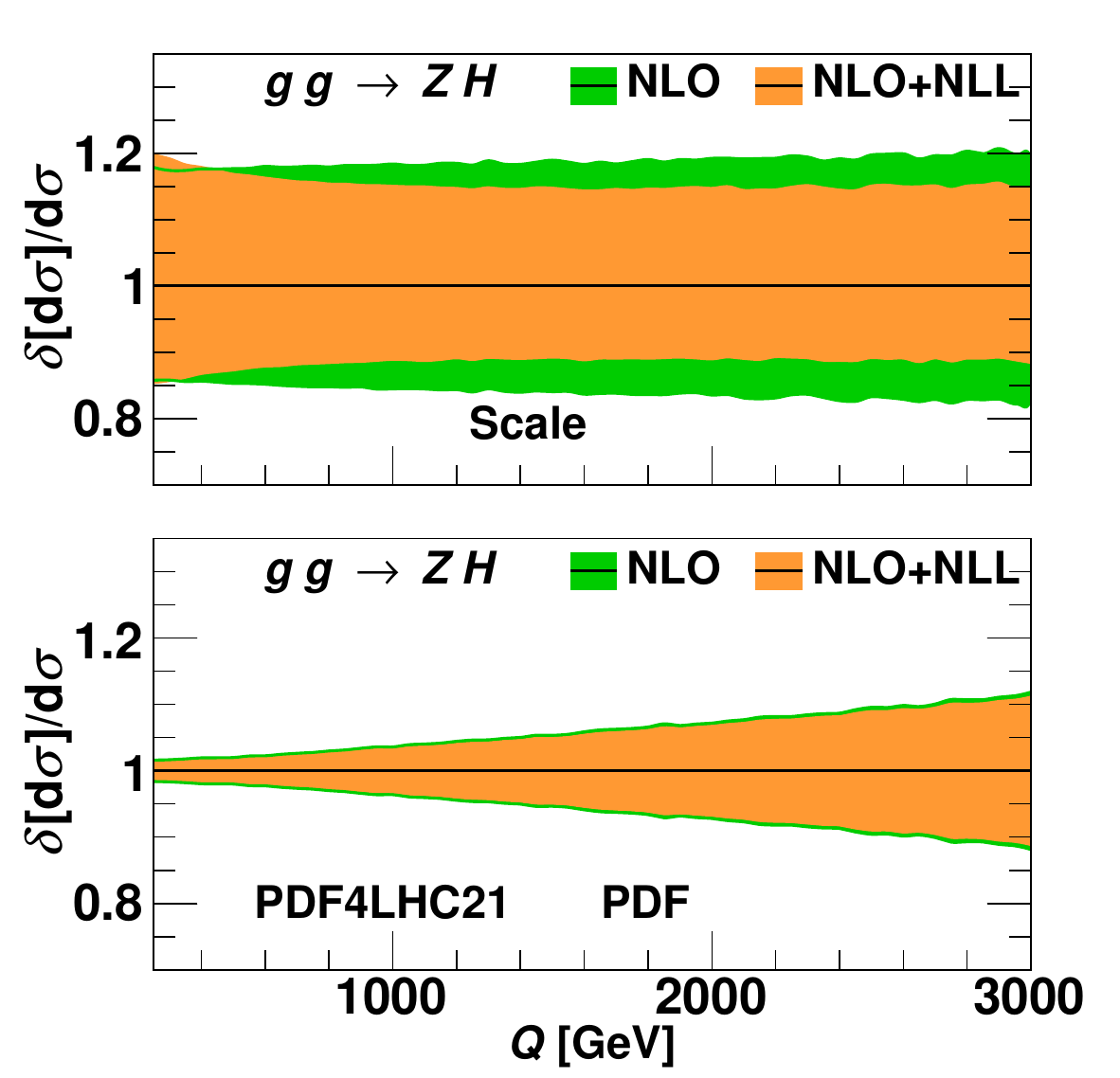}
}
\vspace{-2mm}
\caption{\small{The contribution from $gg$ subprocess is shown for fixed order and resummed cases 	with the corresponding uncertainties.}}
\label{fig:dis_DY_gg}
\end{figure}
The fixed-order results for DY and gluon fusion channel have been computed using the publicly available code \texttt{n3loxs}~\cite{Baglio:2022wzu} and \texttt{vh@nnlo}~\cite{Harlander:2018yio,Brein:2012ne,Altenkamp:2012sx} respectively. 
In the left-top panel of \fig{fig:dis_DY_gg}, we present contributions 
for gluon fusion channels~\cite{Das:2025wbj} along with the seven-point scale uncertainties around central scales $(\mur^c,\muf^c) = (Q,Q)$. To quantify the enhancement due to resummation, we take the ratio 
with respect to NLO which are shown in the left-bottom panel of \fig{fig:dis_DY_gg}. The enhancement due to resummation  in the higher-$Q$ region is about $40\%$ for $gg$ subprocess compared to respective NLO $ggZH$ result. There is also a significant reduction in the scale uncertainty after inclusion of the threshold effects,  by $5.0\%$. 
In the right-top panel of \fig{fig:dis_DY_gg}, we show such a comparison for the scale uncertainty of $gg$ subprocess at NLO and NLO+NLL. In the 
right bottom panel, we present the intrinsic PDF uncertainty for both NLO and NLO+NLL which stays similar for both fixed order and resummed case and amounts to below $2\%$ in the low-invariant mass region.
%
\begin{figure}[ht!]
\centerline{
\includegraphics[scale=0.35]{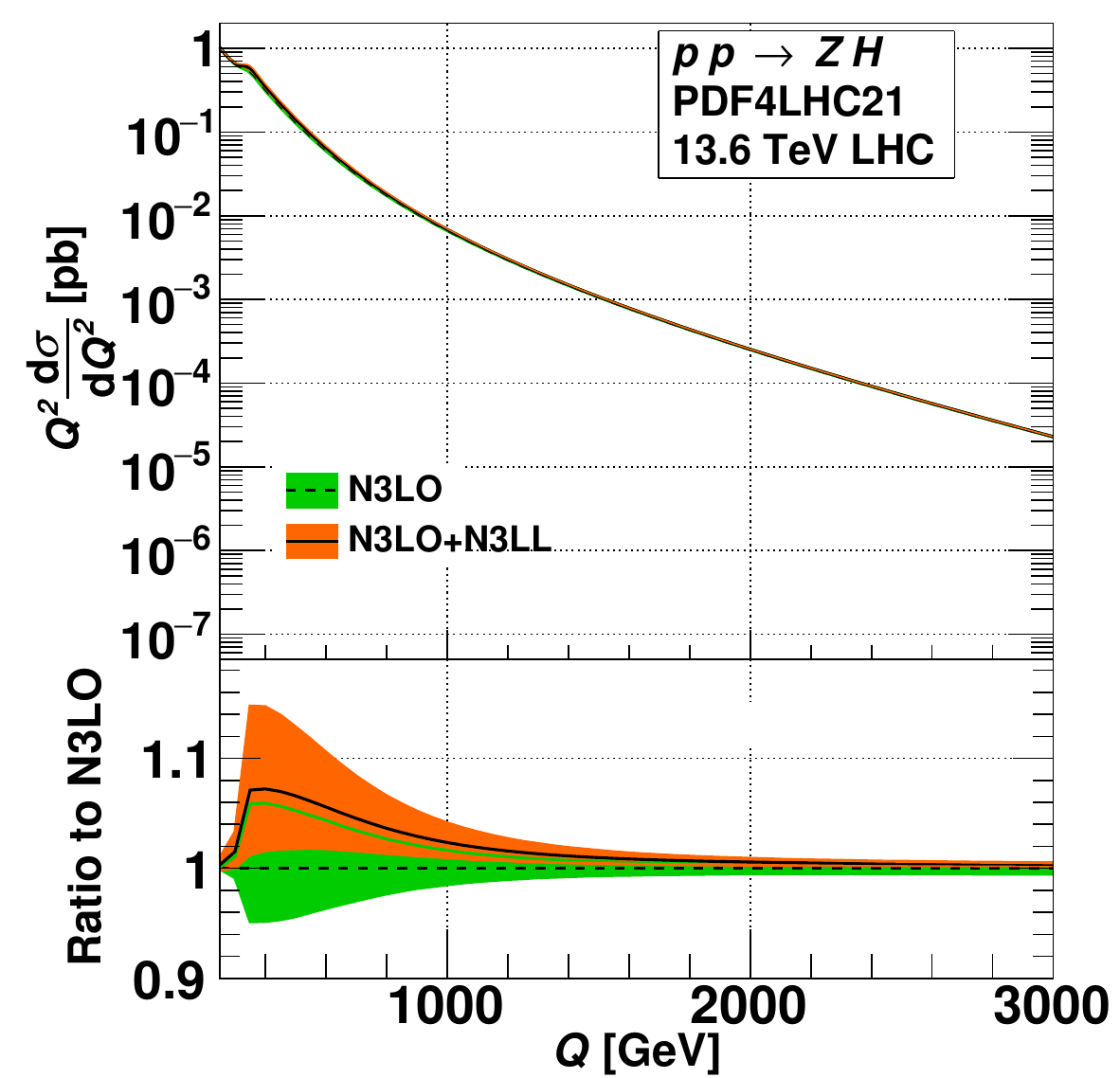}
\includegraphics[scale=0.36]{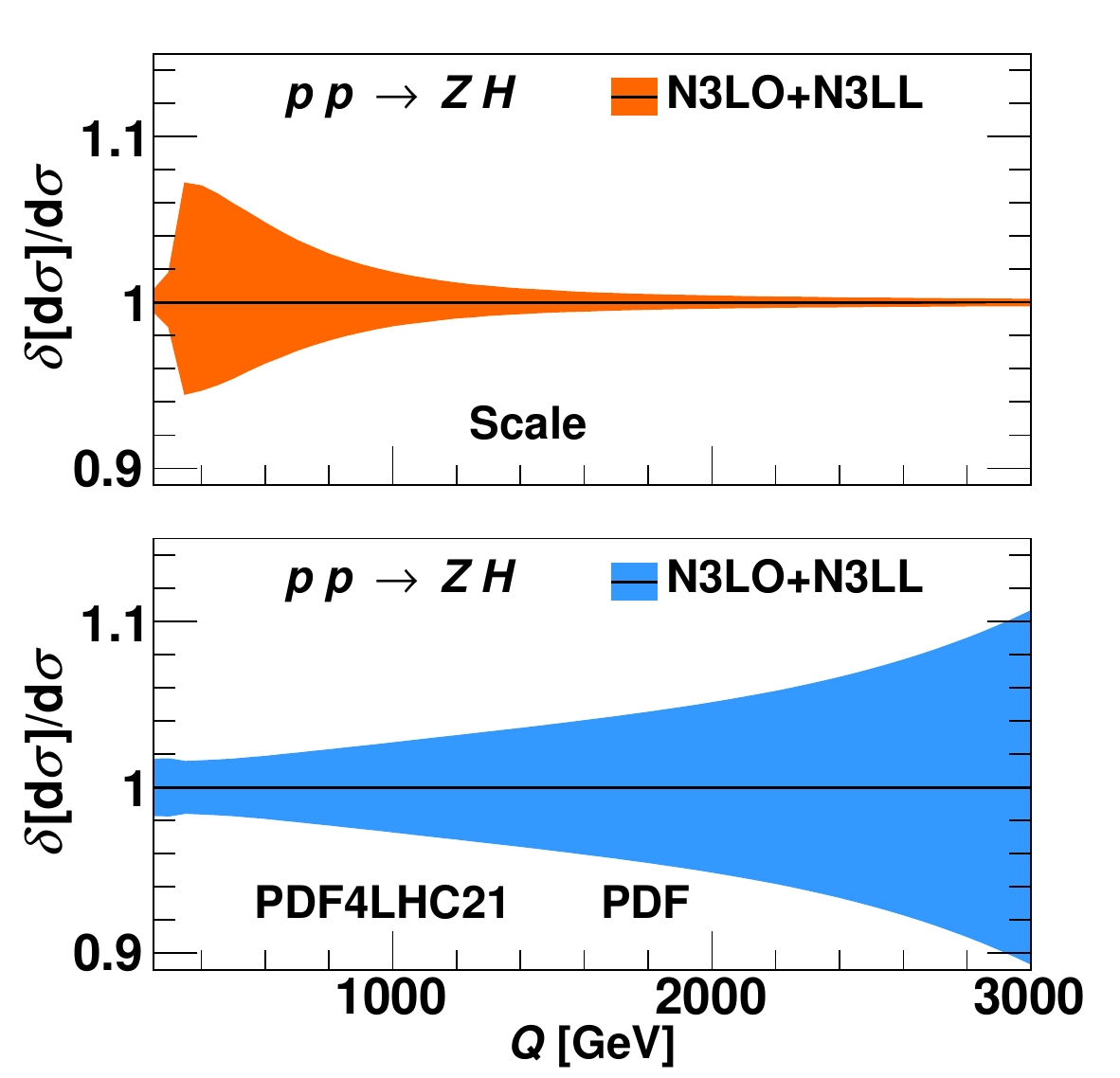}
}
\vspace{-2mm}
\caption{\small{The total contributions for the $pp \to ZH$ process (combination of DY-type and $gg$ subprocess) and the corresponding uncertainties.
    }}
\label{fig:scale_gg}
\end{figure}

For completeness, we combine the gluon fusion and DY-type contributions to obtain the full $pp \to ZH$ results at $\mathcal{O}(\alpha_s^3)$ level. Particularly, for the fixed order case, we combine the N3LO DY contribution to NLO $gg$ contribution to have total N3LO results and in the resummation case, we combine the N3LL DY \cite{Das:2022zie} contribution to NLL $gg$ contribution to have total N3LO+N3LL results. These combined results are shown in \fig{fig:scale_gg}. The right panel of \fig{fig:scale_gg} reflects the corresponding seven-point scale uncertainty and intrinsic PDF uncertainty. We observe that the scale uncertainty remains below $10\%$ around $ZH$ threshold,
while the PDF uncertainty stays below $2\%$ in the same region.
\begin{table}[h!]
	\begin{center}{
	\setlength{\extrarowheight}{5pt}
	\scalebox{1.}{\begin{tabular}{|c|c|c|c|c|c|c|}
						\hline
						Order &
						Central (fb)  &
                        $\Delta$(RESUM) &
						$\delta$(Scale)  &  $\delta$(PDF) & $\delta(\alpha_s)$ & $\delta$(PDF + $\alpha_s$) \\
						\hline
						\hline
						$\sigma^{\text {NLO+NLL}}_{gg}$
						&  $151.3$ & $21.29\%$ & $ \pm 19.4 \%$ &  $\pm 0.7 \%$	& $\pm 2.3 \%$ & $\pm 2.4\% $ \\
                        \hline
						$\sigma^{\text{N}{\rm 3}\text{LO+N}{\rm 3}\text{LL}}_{\rm DY}$
						&  $841.6$ & $0.01\%$ & $\pm 0.6\%$ & $\pm 0.8 \%$
						&  $\pm 0.8 \%$ & $\pm 1.1\% $ \\
                        \hline
						$\sigma^{\text{N}{\rm 3}\text{LO+N}{\rm 3}\text{LL}}_{tot}$
						&  $1004.2 $ & $2.72\%$ & $ \pm 3.0\%$ &  $\pm 0.6 \%$ & $\pm 1.0 \%$ & $\pm 1.2\% $ \\
						\hline
					\end{tabular}
				}}
				\caption{\small{Inclusive resummed cross-sections for $ZH$ are presented (for $gg$, DY type and, total) for $13.6$ TeV with scale, PDF and $\alpha_s$ uncertainties.
                $\sigma^{\text{N}{\rm 3}\text{LO+N}{\rm 3}\text{LL}}_{tot}$ contains both fixed order and  
                resummed results from $gg$ and DY-type as well as contributions from top-loop and bottom annihilation channels at fixed order.}
				}
				\label{tab:tableggZH}
	\end{center}
\end{table}

Finally, in \tab{tab:tableggZH}, we present the total cross-sections including resummation effects from $gg$, DY-type and their combined contribution ($tot$). 
The later contains, not only the $gg$ and DY-type resummation effects, but also all fixed order results including $gg$ at NLO, DY-type at N3LO, contributions from bottom quark annihilation~\cite{Ahmed:2019udm}, top-loop–induced processes~\cite{Brein:2011vx}, where the Higgs boson is radiated from a closed top-quark loop. We present the enhancement due to resummation over fixed order through $\Delta(\text{RESUM}) = (\text{RESUM} -\text{FO})/\text{FO}\times 100\%$. Additionally, we report the associated theoretical uncertainties: the seven-point scale variation, intrinsic PDF uncertainties, and those arising from the strong coupling constant. For the later, we use a $1\sigma$ variation 
$\left(\alpha^{\pm}_S(m_Z) = \alpha^c_S(m_Z) \pm 0.0015\right)$
of strong coupling around its central value $\alpha^c_S(m_Z) = 0.118$. 
The cross-section for $\alpha^{\pm}_S(m_Z)$ has been computed from the 
subsets $41, 42$ according to PDF4LHC recommendation \cite{PDF4LHCWorkingGroup:2022cjn}. The combined PDF+$\alpha_S$ uncertainty is obtained by adding the individual contributions in quadrature. We observe that the largest source of uncertainty arises from the scale variation in gluon fusion channel, which remains sizable at about $19\%$
at NLO+NLL level, indicating the need for further improvements through higher-order computation.

\section{Conclusion}
\label{sec:conclusion}
To summarize, we have studied the impact of soft gluon resummation on 
$Z H$ production at the LHC, focusing in particular on the gluon fusion subprocess. For this channel, we employed the Born-improved NLO framework and matched it with next-to-leading logarithmic (NLL) resummed results. Our analysis shows that soft-virtual (SV) resummation yields an additional enhancement of approximately $20-40\%$  over the fixed-order NLO prediction in the kinematic region considered. Furthermore, the seven-point scale uncertainty is reduced by about $5\%$ compared to the NLO result, with the most notable improvement occurring in the high-$Q$ regime. We have also assessed the uncertainty associated with PDFs in the resummed predictions, finding it to be below $2\%$ in the low invariant mass region.

Finally, to provide experimentally relevant predictions, we combine contributions from all relevant subprocesses—including soft gluon resummation effects in both the $ggZH$ and DY-type channels, and present results for the invariant mass distribution and total production cross section at the $13.6$ TeV LHC.


\paragraph{Funding information}
This research has been supported by the Deutsche Forschungsgemeinschaft (DFG, German Research Foundation) under grant  
396021762 - TRR 257 (\textit{Particle Physics Phenomenology after Higgs discovery.}), the SERB Core Research Grant (CRG) under the project CRG/2021/005270, the Royal Society (URF/R/231031) and the STFC (ST/X003167/1 and ST/X000745/1).
\bibstyle{SciPost_bibstyle}
\bibliography{SciPost_Example_BiBTeX_File.bib}

\end{document}